\documentclass[
aps,     
prl,                    
showpacs,               
superscriptaddress,     
nofootinbib,            
twocolumn,               %
floatfix, nobalancelastpage
]{revtex4}
\usepackage{graphicx}
\usepackage{epsf}

\def\D0bar{\overline D{}^0}

\def\beq{\begin{equation}}
\def\eeq{\end{equation}}
\def\bea{\begin{eqnarray}}
\def\eea{\end{eqnarray}}

\begin{document}

\preprint{\vbox{\hbox{WSU--HEP--0601}}}


\title{\boldmath New Physics contributions to the lifetime difference
in $D^0$-$\overline{D}^0$ mixing}

\author{Eugene Golowich}
\affiliation{Department of Physics,
	University of Massachutetts, Amherst, MA 01003}
\author{Sandip Pakvasa}
\affiliation{Department of Physics and Astronomy,
	University of Hawaii at Manoa, Honolulu, HI 96822}
\author{Alexey A.\ Petrov}
\affiliation{Department of Physics and Astronomy,
	Wayne State University, Detroit, MI 48201}
\affiliation{Michigan Center for Theoretical Physics,
University of Michigan, Ann Arbor, MI 48109}

\begin{abstract}
The first general analysis of New Physics contributions 
to the $D^0$-$\overline{D}^0$ lifetime difference (equivalently 
$\Delta\Gamma_{\rm D}$) is presented.  The extent to which New Physics (NP) 
contributions to $|\Delta C|=1$ processes can produce effects in 
$\Delta\Gamma_{\rm D}$, even if such NP contributions are undetectable 
in the current round of $D^0$ decay experiments, is studied.  
New Physics models which do and do not dominate the lifetime difference 
in the flavor $SU(3)$ limit are identified.  Specific examples 
are provided.
\end{abstract}
\maketitle


Quantum mechanical meson-antimeson oscillations are sensitive
to heavy degrees of freedom which propagate in
the underlying mixing amplitudes. The observation of mixing in 
the $K^0$ and $B_d$ systems thus implied the existence respectively of the 
charm and top quarks before these particles were discovered.
In like manner, by comparing observed meson mixing with
predictions of the Standard Model (SM) modern experimental studies 
have been able to constrain models of New Physics (NP).   

Which system of mixed mesons is likely to produce evidence for NP?
It has become clear from B-factories and the
Tevatron collider that hopes for spectacular NP contributions to
$B_d$ and $B_s$ oscillations have come to naught.  The large SM mixing
succesfully describes all available experimental data.
The only flavor oscillation not yet observed is that of
the charmed meson $D^0$, where SM mixing is very small
and the NP component can stand out.~\cite{Golowich:2005pt}

\begin{figure}[t]
\includegraphics[width=14pc]{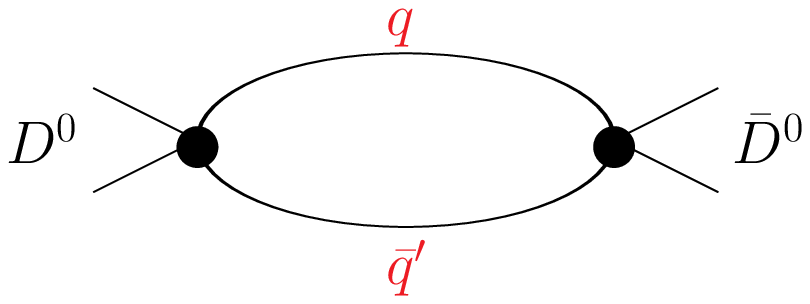}
\caption{Loop diagram for $D^0 \to
{\overline D}^0$.}
\label{fig:loop}
\end{figure}
Charm mixing arises from $|\Delta C|=2$ interactions that generate
off-diagonal terms in the neutral $D$ mass matrix.  To second order, 
the $D^0$-to-${\overline D}^0$ matrix element is 
\begin{eqnarray}\label{M12}
&& \left (M - \frac{i}{2}\, \Gamma\right)_{12} =
  \frac{1}{2 M_{\rm D}}\, \langle \D0bar | 
{\cal H}_w^{\Delta C=-2} | D^0 \rangle \qquad\qquad
\\
&& + \frac{1}{2 M_{\rm D}}~\, \sum_n {\langle \D0bar | {\cal H}_w^{\Delta
  C=-1} | n \rangle\, \langle n | {\cal H}_w^{\Delta C=-1} 
| D^0 \rangle \over M_{\rm D}-E_n+i\epsilon}\,
\nonumber
\end{eqnarray}
where ${\cal H}_w^{|\Delta C|=1,2}$ is the effective
$|\Delta C|=1,2$ hamiltonian.  The most natural place for NP 
to affect mixing amplitudes is in the $|\Delta C|=2$ piece, which 
corresponds to a local interaction at the charm quark mass scale.  
This local interaction cannot, however, affect 
$\Delta \Gamma_{\rm D}$ because it does not have an absorptive
part. 

Let us introduce standard notation for 
$\Delta \Gamma_{\rm D}$ and $\Delta M_{\rm D}$ 
by employing the dimensionless forms, 
\bea
y = {\Delta \Gamma_{\rm D} \over 2\Gamma_{\rm D}} \ ,
\qquad  \qquad 
x = {\Delta M_{\rm D} \over \Gamma_{\rm D}} \ \ . 
\eea
Given CP-conservation, we can express $y$ 
as an absorptive part of Eq.~(\ref{M12}),
\begin{equation}\label{y1}
y=\frac{1}{\Gamma_{\rm D}}\sum_n \rho_n
\langle  \overline{D}^0| {\cal H}_w^{\Delta C=-1}| n \rangle
\langle n | {\cal H}_w^{\Delta C=-1}| D^0 \rangle,
\end{equation}
where $\rho_n$ is a phase space function that corresponds to 
charmless intermediate state $n$.  This relation shows that 
$\Delta \Gamma_{\rm D}$ is driven by 
transitions $D^0, {\overline D}^0 \to n$, 
{\it i.e.} physics of the $|\Delta C|=1$ sector.  
It turns out that experimentally observed $D^0$ decays 
agree reasonably well with SM estimates~\cite{Buccella:1994nf}.  
To date, no clear signals of NP have been 
observed~\cite{Bergmann:1999pm}.
As such, it is currently accepted that $\Delta\Gamma_{\rm D}$ 
is dominated by the SM contribution.  
In this Letter, we show that this is not necessarily
so and consider several NP models to illustrate our point.  

Consider a $D^0$ decay amplitude which includes
a small NP contribution, $A[D^0 \to n]=A_n^{\rm (SM)} + A_n^{\rm (NP)}$.
%
%
Here, $A_n^{\rm (NP)}$ is assumed to be smaller than the current experimental
uncertainties on those decay rates. Then it is a good approximation 
to write Eq.~(\ref{y1}) in the form
\begin{eqnarray}\label{schematic}
y &\simeq& \sum_n \frac{\rho_n}{\Gamma_{\rm D}} 
A_n^{\rm (SM)} \bar A_n^{\rm (SM)}
+ 2\sum_n \frac{\rho_n}{\Gamma_{\rm D}}
A_n^{\rm (NP)} \bar A_n^{\rm (SM)} \ \ . 
\label{approx}
\end{eqnarray}
The first term in this equation corresponds to 
SM interactions at both vertices in Fig.~\ref{fig:loop},
whereas for the second term, there is one SM vertex 
and one NP vertex.

The SM contribution to $y$ is known to vanish
in the limit of exact flavor $SU(3)$~\cite{Donoghue:1985hh}.
Moreover as was shown in~\cite{FalkEtAl},
the first order correction is also absent, so the SM 
contribution arises only as a {\it second} order effect.
Thus, those NP contributions which do not vanish in the flavor
$SU(3)$ limit must determine the lifetime difference there,
even if their contibutions are tiny in the
individual decay amplitudes. The same reasoning 
can be applied to $x$ since a dispersion relation relates $x$ to
$y$~\cite{Falk:2004wg}.

Of course, flavor $SU(3)$ symmetry is broken in the real world.
Just how large this effect is on 
$D^0$-$\overline{D}^0$ mixing in the SM 
is controversial, with estimates for 
$y$ ranging from a percent~\cite{Falk:2004wg} to orders of 
magnitude smaller~\cite{Buccella:1994nf,Petrov:2003un}. 
The current experimental bounds on $y$ and $x$ are~\cite{pdg}
\begin{eqnarray*}
y  < 0.008 \pm 0.005 \ , \quad \quad 
 x < 0.029 \ \
(95\%~ {\rm C.L.}) \nonumber\ \ . 
\end{eqnarray*}
A NP $|\Delta C|=1$ interaction can have a measureable effect on 
the value of $y$ (and of $x$) if the true SM value for $y$ does not 
near the top of the range of predictions. 
We shall assume that this is the case.

Using the completeness relation and Eq.~(\ref{y1}), the NP
contribution to the $D^0$-${\overline D}^0$ lifetime difference 
becomes 
\begin{eqnarray}\label{gammaope}
y \ &=& \  \frac{2}{M_{\rm D} \Gamma_{\rm D}}\, \langle \D0bar |
    {\rm Im}\, {\cal T} | D^0 \rangle \ \ ,
\\
{\cal T} \ &=& \
    \,i\! \int\! {\rm d}^4 x\, T \left(
    {\cal H}^{\Delta C=-1}_{SM} (x)\, {\cal H}^{\Delta C=-1}_{NP}(0) 
\right) \ \ .\nonumber
\end{eqnarray}
We represent the NP $\Delta C=-1$ hamiltonian as\footnote{Throughout 
this paper, we reserve indices $i,j,k,\ell$ for color.}
\bea\label{HamNP}
&& {\cal H}^{\Delta C=-1}_{NP} =
\sum_{q,q'} \ D_{qq'}
\left[\overline {\cal C}_1(\mu) Q_1 +
\overline {\cal C}_2 (\mu) Q_2 \right]\ ,
\\
&& Q_1 = \overline{u}_i \overline\Gamma_1 q_j' ~
\overline{q}_j \overline\Gamma_2 c_i \ , 
\ \
Q_2 = \overline{u}_i \overline\Gamma_1 q_i' ~
\overline{q}_j \overline\Gamma_2 c_j\ , 
\nonumber
\eea
where the spin matrices 
$\overline\Gamma_{1,2}$ can have arbitrary Dirac structure, 
$\overline {\cal C}_{1,2}(\mu)$ are
Wilson coefficients evaluated at
energy scale $\mu$ and the flavor sums on $q,q'$ 
extend over the $d,s$ quarks. We shall expand the time-ordered product
of Eq.~(\ref{gammaope}) in an operator product expansion (OPE), 
{\it i.e.} in terms of local operators of 
increasing dimension~\cite{ShD}.

The leading term in the OPE is simply that depicted in 
Fig.~\ref{fig:loop}.  For a generic NP interaction, we 
calculate that 
\bea\label{yNP}
y \ &=& - \ \frac{4\sqrt{2} G_F}{M_{\rm D} \Gamma_{\rm D}} \
\sum_{q,q'} {\bf V}_{cq'}^* {\bf V}_{uq} D_{qq'}
\left(K_1 \delta_{ik}\delta_{j\ell} \right.
\nonumber \\
&+& \
\left. K_2 \delta_{i\ell}\delta_{jk} \right)
\sum_{\alpha=1}^5 \ I_\alpha (x,x') \ \langle
\overline{D}^0| \ {\cal O}_\alpha^{ijk\ell} \ | D^0 \rangle, \ \
\eea
where $\{K_\alpha\}$ are combinations of Wilson coefficients, 
\begin{eqnarray}
K_1= \left({\cal C}_1 \overline{\cal C}_1 N_c
+ \left({\cal C}_1 \overline{\cal C}_2 + \overline{\cal C}_1
{\cal C}_2 \right)\right), \
K_2 = {\cal C}_2 \overline{\cal C}_2 \ \  
\end{eqnarray}
with the number of colors $N_c = 3$.  
The operators $\{ {\cal O}_i^{ijk\ell}\}$ in Eq.~(\ref{yNP}) are defined as
\bea
{\cal O}_1^{ijk\ell} \ &=& \
\overline{u}_k \Gamma_\mu \gamma_\nu \overline\Gamma_2 c_j ~
\overline{u}_\ell \overline\Gamma_1 \gamma^\nu \Gamma^\mu c_i \,
\nonumber \\
{\cal O}_2^{ijk\ell} \ &=& \ \overline{u}_k \Gamma_\mu 
\hspace{-0.14cm} \not p_c
\overline\Gamma_2 c_j ~
\overline{u}_\ell \overline\Gamma_1 \not p_c \Gamma^\mu c_i \,
\nonumber \\
{\cal O}_3^{ijk\ell} \ &=& \ \overline{u}_k \Gamma_\mu \overline\Gamma_2 c_j ~
\overline{u}_\ell \overline\Gamma_1 
\hspace{-0.14cm} \not p_c \Gamma^\mu c_i \,
\label{ops} \\
{\cal O}_4^{ijk\ell} \ &=& \ \overline{u}_k \Gamma_\mu 
\hspace{-0.14cm} \not p_c
\overline\Gamma_2 c_j ~
\overline{u}_\ell \overline\Gamma_1 \Gamma^\mu c_i \,
\nonumber \\
{\cal O}_5^{ijk\ell} \ &=& \ \overline{u}_k \Gamma_\mu \overline\Gamma_2 c_j ~
\overline{u}_\ell \overline\Gamma_1 \Gamma^\mu c_i \ \ , \nonumber
\eea
where $p_c$ is the charm-quark momentum operator, 
$\Gamma_\mu \equiv \gamma_\mu P_L$, 
$P_L \equiv (1 + \gamma_5)/2$ and later we also use 
$P_R \equiv (1 - \gamma_5)/2$. The coefficients $I_\alpha(x,x')$ in 
Eq.~(\ref{yNP}) are 
\bea
I_1(x,x')&=&-\frac{k^* m_c}{48\pi}\left[
1- 2\left(x+x'\right)+\left(x-x'\right)^2
\right]
\nonumber \\
I_2(x,x')&=&-\frac{k^*}{24\pi m_c}
\left[
1+\left(x+x'\right)-2\left(x-x'\right)^2
\right]
\nonumber \\
I_3(x,x')&=&\frac{k^*}{8\pi}  \sqrt{x}
\left(1+x'-x\right)  \label{Is}
\\
I_4(x,x')&=&-\frac{k^*}{8\pi} \sqrt{x'}
\left(1-x'+x\right)
\nonumber \\
I_5(x,x')&=&\frac{k^* m_c}{4\pi} \sqrt{x x'} \ \ ,
\nonumber
\eea
where $k^* \equiv (m_c/2)[1-2 (x + x') + (x - x')^2]^{1/2}$ with 
$x\equiv m_q^2/m_c^2$ and $x'\equiv m_{q'}^2/m_c^2$.

Equations (\ref{yNP})-(\ref{Is}) represent the basic formulas 
in our analysis. Hereafter, we take $m_d = 0$ and express 
results in terms of 
the Wolfenstein parameter $\lambda \equiv {\bf V}_{us} 
= - {\bf V}_{cd} \simeq 0.22$ and $x_s \equiv m_s^2/m_c^2 
\simeq 0.006$.  All our results are presented to leading order 
in $x_s$. Finally, predictions using NP masses and couplings 
other than the ones assumed here can be obtained via simple scaling.  

\vspace{0.1cm}

The SM hamiltonian ${\cal H}^{\Delta C=-1}_{\rm SM}$
is recovered in Eq.~(\ref{HamNP})
by setting $D_{qq'}=-({G_F}/{\sqrt{2}}) {\bf V}_{cq}^*{\bf V}_{uq'}$,
$\overline {\cal C}_i \to {\cal C}_i$, 
$\overline\Gamma_{1,2} \to \Gamma_\mu$, and the known SM result 
$y_{\rm SM}^{\rm (LO)}$ easily follows, 
\bea
y_{\rm SM}^{\rm (LO)} &=& 
\frac{G_F^2 m_c^2 \lambda^2 x_s^3}{2 \pi M_{\rm D} \Gamma_{\rm D}} 
\left(K_2 - K_1\right) \left[\langle Q \rangle +
4 \langle Q_S \rangle\right]\ ,
\nonumber \\
\langle Q \rangle &=& \langle \overline D^0 | 
\overline{u}_{i} \Gamma_\mu c_{i}
\ \overline{u}_{j} \Gamma^\mu c_{j} | D^0 \rangle \ ,
\label{smeq} \\
\langle Q_S \rangle &=& \langle \overline D^0 | 
\overline{u}_{i} P_R c_{i}
\ \overline{u}_{j} P_R c_{j} | D^0 \rangle\ \ .
\nonumber
\eea
Note that this contribution is suppressd by six powers of $m_s$ and is 
therefore tiny, $y_{\rm SM}^{\rm (LO)} \sim
10^{-8}$~\cite{Golowich:2005pt}.  This is because the GIM mechanism 
requires four chirality flips (strange quark mass insertions) for 
the intermediate quarks plus additional helicity flip due to the 
pseudoscalar initial state.~\cite{FalkEtAl}.

\vspace{0.2cm}

\centerline{\bf Examples of New Physics Models}

\vspace{0.1cm}

In what follows, we distinguish between NP models which vanish 
in the limit of SU(3) flavor symmetry from those which do not.

{\bf Nonzero SU(3) Limit}: For NP models with flavor-dependent
couplings $D_{qq'}$, it is possible to obtain contributions 
that are nonzero in the flavor $SU(3)$ limit. For these, the main 
contibution will come from the operators ${\cal O}_{1,2}$ (as 
${\cal O}_{3,4,5}$ are suppressed by powers of $m_s/m_c$). 
The two most common scenarios involve (V-A)$\otimes$(V-A) and
(S-P)$\otimes$(S+P) couplings.

As an example, consider a NP model whose low energy effective 
hamiltonian is represented by a four-fermion operator with vectorial 
left-handed interactions,
{\it i.e.}
$\overline\Gamma_1=\overline\Gamma_2= \gamma_\mu P_L$
and $D_{qq'}=\lambda_{qq'}/\Lambda^2$, where $\Lambda$ is the 
NP mass scale. We find
\beq\label{yNPvector}
y_{\rm VLH} = \frac{C \widetilde{\lambda}}{\Lambda^2}
\left[ 
{ K_1 + 2 K_2 \over 2} \langle Q \rangle
+ \left(K_2-K_1\right) \langle Q_S \rangle
\right], \
\eeq
where $C\equiv \sqrt{2} G_F m_c^2/(3 \pi M_{\rm D} \Gamma_{\rm D})$ and 
\beq
\widetilde{\lambda}=
\lambda_{sd} - \lambda \left(\lambda_{dd}-\lambda_{ss}\right)-
\lambda^2 \left(\lambda_{ds} + \lambda_{sd}\right) 
\label{lamb}
\eeq
is the combination of NP couplings to the $s,d$ quarks. 

It follows from Eq.~(\ref{lamb}) that 
if all NP couplings $\lambda_{qq'}$ are of the same order, 
$y_{\rm VLH}$ is nonzero in the flavor $SU(3)$ limit. The result for a 
penguin-like NP contribution can be obtained if one sets
$\lambda_{sd}=\lambda_{ds}=0$. In this case the same conclusion holds if
$\lambda_{dd}\neq\lambda_{ss}$ (which is however not easy
to arrange) unless a (generally tiny) $|A_n^{\rm (NP)}|^2$ term 
is also included in Eq.~(\ref{schematic}). In what follows, we will be 
neglecting QCD running of the local operators
generated by the NP interaction ({\it i.e.} $\overline{\cal C}_1=0$ 
and $\overline{\cal C}_2=1$).

{\it Models with extra vector-like quarks}: Consider a model of 
the above type which extends the SM by including new singlet quarks
in a vector-like representation~\cite{bpr}. In this instance, the 
$Z$-boson has additional flavor-changing couplings. 
For example, assume both up-type and down-type exotic quarks 
$U_{a,i}$, $D_{a,i}$ are present 
(indices $a,i$ denote flavor and color respectively).  Then the 
flavor-changing couplings are described by
\bea
{\cal L}_{xQ} \ &=& \ -\frac{g}{2\cos\theta_{\rm W}} \
J^{\rm (NC)}_\mu Z^\mu + \mbox{h.c.} \ \ ,
\label{xQ} \\
J^{\rm (NC)}_\mu \ &=& \ {\bf U}_{ab}^{\rm (u)} \ 
\overline{U}_{a,i} \Gamma_\mu  U_{b,i}
+ {\bf U}_{ab}^{\rm (d)} \ \overline{D}_{a,i} \Gamma_\mu D_{b,i} \ ,
\nonumber 
\eea
with flavor-changing couplings in both up and down sectors. 
The lifetime difference for this model can be obtained from 
Eq.~(\ref{yNPvector}) by substituting 
$\lambda_{sd}={\bf U}_{cu}^{\rm (u)} 
{\bf U}_{sd}^{{\rm (d)}\dagger}$, 
$\lambda_{ds}=\lambda_{dd}=\lambda_{ss}=0$ 
and $\Lambda=\sqrt{2/G_F}$. This model is well-constrained from 
measurements of the mass differences
in $K\overline{K}$ and $D\overline{D}$ mixing. 
For ${\bf U}_{cu}^{\rm (u)} \sim 10^{-3}$
and ${\bf U}_{sd}^{\rm (d)} \sim 10^{-4}$ we get 
$y \sim 10^{-8}$, of the same order of
magnitude as $y_{\rm SM}^{\rm (LO)}$ above. It is 
worth noting that the little Higgs
models, which have similar low-energy 
signatures, are not constrained by the
measurements of lifetime difference, 
as they do not have flavor-changing couplings
for the down quark sector (flavor-conserving 
contributions cancel out in $y$).

{\it SUSY without R-parity (slepton exchange):} Another example 
of a contribution which survives in the flavor $SU(3)$ limit is 
SUSY without R-parity~\cite{bt}. In this model, there are flavor-changing 
interactions of sleptons that can be obtained from the lagrangian
\bea
{\cal L}_{\not{R}} \ = \  \lambda_{ijk}'L_i Q_j D^c_k \ \ ,
\eea
as well as the interactions mediated by squarks discussed below.
The slepton-mediated interaction is not suppressed in the flavor SU(3) 
limit and leads to 
\bea\label{RPVSUSY}
y_{\not{R}}=
\frac{C' \widetilde{\lambda}}{M_{\widetilde\ell}^2}
\left[
\left( C_2 - 2 C_1\right) \langle Q^\prime\rangle +
\left( C_1 - 2 C_2\right) \langle \widetilde Q^\prime\rangle
\right],
\eea
where $C'=-G_F m_c^2/(6 \sqrt{2} \pi M_{\rm D} \Gamma_{\rm D})$, 
$M_{\widetilde\ell}$ is a slepton mass,  
$\widetilde\lambda$ is given by 
Eq.~(\ref{lamb}) with 
$\lambda_{sd}=\lambda'_{i12}\lambda'_{i21} \le 1 \times 10^{-9}$, 
$\lambda_{ss}=\lambda'_{i11}\lambda'_{i21} \le 5 \times 10^{-5}$,
$\lambda_{dd}=\lambda'_{i21}\lambda'_{i22} \le 5 \times 10^{-5}$, 
$\lambda_{ds}=\lambda'_{i11}\lambda'_{i22} \le 5 \times 10^{-2}$~\cite{bt},
and $\langle Q^\prime\rangle$ is
\bea\label{Qprime}
\langle Q' \rangle  \ &=& \ \langle \overline D^0 |  \overline{u}_{i} 
\gamma_\mu P_L c_{i}
\ \overline{u}_{j} \gamma^\mu P_R c_{j} | D^0 \rangle \ \ .
\eea
Operators with a tilde are obtained by swapping color indices in the 
charm quark operators. Using factorization to estimate matrix elements 
of the above operators and taking for definiteness 
$M_{\widetilde\ell} = 100$~GeV, we arrive at
$y_{\not{R}} \simeq -3.7\%$. The contribution due 
to squark exchange vanishes in the flavor SU(3) limit and is given below.

{\bf Zero SU(3) Limit}: There are several reasons that some NP models 
vanish in the flavor $SU(3)$ limit.  First, the structure of the
NP interaction might simply mimic the one of the SM. Effects like that 
can occur in some models with extra space dimensions. Second,
the chiral structure of a low-energy effective lagrangian in 
a particular NP model could be such that the leading, 
mass-independent contribution vanishes exactly, 
as in a left-right model (LRM). Third, the NP coupling might explicitly
depend on the quark mass, as in a model with multiple
Higgs doublets. There, the charged Higgs couplings explicitly 
depend on quark mass. However, most of these models feature 
second order $SU(3)$-breaking already at leading order in the $1/m_c$ 
expansion. This should be contrasted with the SM, where the 
leading order is suppressed by six
powers of $m_s$ and the second order only appears as a 
$1/m_c^6$-order correction.

{\it Left-right models}: Left-right models (LRM) provide new 
tree-level contributions mediated by right-handed ($W^{\rm (R)}$) 
bosons~\cite{mp}. 
The relevant effective lagrangian is
\beq
{\cal L}_{\rm LR}=-\frac{g_{\rm R}}{\sqrt{2}} \ {\bf V}_{ab}^{\rm (R)} \
\overline{u}_{a,i} \gamma^\mu P_R d_{b,i}~ W^{\rm (R)}_\mu +
\mbox{h.c.}  \ \ , 
\eeq
where ${\bf V}^{\rm (R)}_{ik}$ are the coefficients of the 
right-handed CKM matrix.
This leads to a local $\Delta C=-1$ hamiltonian as in 
Eq.~(\ref{HamNP}) with $\overline\Gamma_1=\overline\Gamma_2 
= \gamma_\mu P_R$.  Since 
current experimental limits allow $W^{\rm (R)}$ masses as low as 
a TeV~\cite{pdg}, 
a sizable contribution to $y$ is quite possible. Using Eq.~(\ref{yNP}),
we obtain 
\bea
& & y_{\rm LR} = - C_{\rm LR} {\bf V}^{\rm (R)}_{cs} 
{\bf V}^{\rm (R)*}_{us} 
\left[ {\cal C}_1 \langle Q' \rangle
+ {\cal C}_2 \langle \tilde Q' \rangle 
\right] \ \ , \label{yLRM}
\eea
where $C_{\rm LR} \equiv \lambda G_F^{\rm (R)} 
G_F m_c^2 x_s/(\pi M_{\rm D}\Gamma_{\rm D})$, 
$G_F^{\rm (R)}/\sqrt{2} \equiv g^2_{\rm R}/8 M^2_{W_{\rm R}}$, 
${\cal C}_{1,2}$ are the SM Wilson coefficients and the operators
appearing in Eq.~(\ref{yLRM}) are given in Eq.~(\ref{Qprime}).
Using~\cite{pdg}, we obtain numerical values for two possible 
realizations: (i) 'Manifest LR' (${\bf V}^{\rm (L)} = {\bf V}^{\rm (R)}$) 
gives $y_{\rm LR} = -4.8 \cdot 10^{-6}$ with $M_{W_{\rm R}} = 1.6$~TeV 
and (ii) 'Nonmanifest LR' (${\bf V}^{\rm (R)}_{ij} \sim 1$) gives 
$y_{\rm LR} = -8.8 \cdot 10^{-5}$ with $M_{W_{\rm R}} = 0.8$~TeV.
In both cases we take $g_{\rm R} = g_{\rm L}$.
  
{\it Multi-Higgs models}: A popular realization of this type 
is the two Higgs doublet model (2HDM) with natural flavor conservation. 
This model provides new tree-level
contributions mediated by charged Higgs bosons and 
leads to the local four fermion interaction~\cite{Golowich:1978nh}
\bea
{\cal H}_{\rm ChH}^{\Delta C=-1} = -
\frac{\sqrt{2} G_F}{M_{\rm H}^2}  ~
\overline{u}_i \overline\Gamma_1 q_i' ~
\overline{q}_j \overline\Gamma_2 c_j \ \ ,
\eea
where the vertices $\overline\Gamma_{1,2}$ are
\bea
\overline\Gamma_1 &=& m_{q'} \cot\beta ~V_{uq'} ~
 P_R - m_u \tan\beta ~V_{uq'} ~ P_L \ , 
\nonumber \\
\overline\Gamma_2 &=& m_q \cot\beta ~V_{cq}^* ~P_L 
- m_c  \tan\beta ~ V_{cq}^* ~ P_R \ \ .
\label{g1g2}
\eea
There are four possible contributions involving the 
various terms in $\overline\Gamma_{1,2}$.  However, three 
of these, including the potentially large $\tan^2\beta$ term, 
vanish for assorted reasons ({\it e.g.} flavor cancellation, 
zero matrix element).  This leaves 
\bea
& & y_{\rm ChH} = {\lambda^2 G_F^2 m_c^4 x_s^{3/2}  \over 
\pi M_{\rm D} \Gamma_{\rm D} M_{\rm H}^2} ~\cot^2 \beta  
\left[ {\cal C}_1 + {\cal C}_2 \right] 
\langle {\cal Q} \rangle  \ ,
\label{chh} 
\eea
where $\langle {\cal Q} \rangle$ is as in Eq.~(\ref{smeq}).
Assuming values $M_{\rm H} = 85$~GeV and $\cot\beta = 0.05$, 
consistent with constraints obtained from the observation 
of $B \to \tau \nu_\tau$~\cite{pdg}, we obtain $y_{\rm ChH} 
\simeq 2\cdot 10^{-10}$.  

{\it SUSY without R-parity (squark exchange):} The baryon-number violating 
squark exchanges arise from the lagrangian~\cite{bt}  
\bea
{\cal L}_{\not{R}} \ = \  \lambda_{ijk}''U_i^c D_j^c D^c_k \ \ .
\eea
This interaction has the same Dirac structure
as the LRM discussed earlier and leads to 
\bea\label{RPVSUSY2}
y'_{\not{R}}=
- x_s C'' \frac{\lambda_{22k}''\lambda_{12k}''}{M_{\widetilde{sq}}^2}
\left[
{\cal C}_2 \langle Q^\prime\rangle +
{\cal C}_1 N_c \langle \widetilde Q^\prime\rangle
\right] \ \ ,
\eea
where $C''= G_F \lambda/(2 m_D \pi \Gamma)$, 
$M_{\widetilde sq}$ is a squark mass and the matrix elements 
$\langle Q' \rangle$, $\langle \widetilde Q'\rangle $ are given
earlier. Using factorization for the matrix elements, 
$\lambda_{22k}''\lambda_{12k}'' \sim 3\cdot 10^{-4}$~\cite{bm}, 
and taking $M_{\widetilde{sq}} = 100$~GeV,  
we arrive at the result $y'_{\not{R}}\simeq 6.4\cdot 10^{-6}$. 

\vspace{0.5cm} 
In conclusion, we have explored how NP contributions can influence 
the lifetime difference $y$ in the charm system. We argued that 
the NP signal is dominant in the formal flavor $SU(3)$ limit. 
We also showed that, for some NP models, it is possible that small 
NP contributions to $|\Delta C|=1$ processes produce substantial 
effects in the $D^0\overline{D}^0$ lifetime difference, even if 
such contributions are currently undetectable in the experimental 
analyses of charmed meson decays. Coupled with a known difficulty in 
computing SM contributions to D-meson decay amplitudes, it might be 
advantageous to use experimental constraints on $y$ in order to test 
various NP scenarios due to better theoretical 
control over the NP contribution and SU(3) suppression of the SM 
amplitude.

We thank X. Tata and J. Hewett for clarifying discussions.  
The work of E.G. was supported in part by the U.S.\ National Science
Foundation under Grants PHY-0244801 and PHY-0555304. S.P. was
supported by the U.S.\ Department of Energy under Contract DE-FG02-04ER41291.
A.P.~was supported in part by the U.S.\ National Science Foundation
CAREER Award PHY--0547794, and by the U.S.\ Department of Energy under Contract
DE-FG02-96ER41005.


\end{document}